\documentstyle[preprint,prl,aps]{revtex}
\tightenlines
\begin{document}
\draft

\title{Parity violation in reactions with B- and Li-nuclei using polarized
thermal neutrons}

\author
{V. A. Vesna$^1$, Yu. M. Gledenov$^2$, P.V.Lebedev-Stepanov$^3$, 
I.S.Okunev$^1$,
\\ Yu. P. Popov$^2$,
A.V.Sinyakov$^3$, E.V.Shul'gina$^1$,  
Yu.M.Tchuvil'sky$^3$}
  \maketitle
  \begin{center}
$^1${\it Petersburg Nuclear Physics Institute, Gatchina 188350, Russia}\\
$^{2}${\it Joint Institute for Nuclear Research, Dubna 141980, Russia}\\
$^{3}${\it Moscow State University, Moscow 119899, Russia}
\end{center}

\begin{abstract}
Parity violation in the reactions $^6$Li(n,$\alpha$)$^3$H,
$^{10}$B(n,$\alpha$)$^7$Li  and $^{10}$B(n,$\alpha$)$^7$Li$^* \rightarrow$ 
$\gamma + ^7$Li(g.s) is discussed. It appears that the effects
(neutral currents contribution in particular) are
measurable due to the large values of the neutron cross sections. This  holds
promise for the theoretical extraction of the constants of the weak
NN potential because of the relative simplicity of the investigated systems.
\end{abstract}

\clearpage

The fact that the list of measured parity violation (PV)
processes in nuclei accessible for extracting  the
weak NN interaction constants is short sends investigators in search for
new regions.
Moreover, the current data on PV in nuclei are, for the most part,
in the region of heavy or middle (A$>20$) atomic weight nuclei.
The possibility of extracting the constants of the parity violating NN
Hamiltonian
using such data is the subject of sharp controversy.
Almost the same is true for most experimental results in the
light nuclei region (A$\sim$ 14-20).
Only a few measured examples are suitable here for 
extracting the constants of the P-odd part of the NN-potential.
The theoretical situation seems to be better for few-nucleon systems
(A$\leq 4$) and lightest nuclei (4$<$ A$\leq 10$).
However, in these cases another problem arises. This problem 
is that almost all measured examples of nuclear P-odd processes for A$>$4
are connected with the existence of parity mixing doublets (PMD) \cite {AH}.
At the same time, the spectra of the majority of lightest nuclei
do not contain such doublets.
As a result, P-odd  effects in these nuclei are not enhanced. In addition,
the theory of P-odd nuclear effects in the absence of PMD is
developed for few special cases only.

The idea to use the discussed reactions to measure the PV effect together
with some estimates are reported in Ref. \cite {ld}.
Requirements for the necessary experimental precision turn out to be
very strict. So, the sole subject of intensive investigations
in the 4$<$A$\leq 10$ region was the parity
forbidden $\alpha $-decay of the 3.56 MeV level in the $^6$Li nucleus. 
However, PMD is
also absent in it and the effect turned out to be
extremely small according to various calculations \cite{rb,bdktt}.

In the present work, we demonstrate the results of both experimental and
theoretical efforts to do the investigation of parity violation in the
region of lightest nuclei. Particular attention has been given to the
discussion of P-odd effect in $\gamma$-transition in $^7$Li-nucleus where
there is a definite possibility to extract PV constants.

A joint PNPI-JINR group carried out experimental research of the P-odd
asymmetries $ W^{\alpha}_{PV}(\theta)\sim
1+a^{\alpha}_{PV}({\bf s}_n \cdot {\bf k}_{\alpha})$
and $ W^{\gamma}_{PV}(\theta ) \sim
1+a^{\gamma}_{PV}({\bf s}_n {\bf k}_{\gamma})$
in the reactions $^{10}$B(n,$\alpha_{0,1}$)$^7$Li and
$^6$Li(n,$\alpha$)$^3$H.
The measurements were performed in the vertical beam of
polarized thermal neutrons from the water-water reactor (WWR-M)
in PNPI. The neutron beam had dimensions of 100x30 mm$^2$ with an  
intensity of $3\cdot 10^{10}$ neutrons/s.
The neutron polarization averaged over the beam
cross section amounted to P=0.8. In all discussed reactions,
we observed comparatively large left-right asymmetries of the form
$ W_{LR} \sim 1+a_{LR}({\bf s}_{n} [{\bf k}_{\alpha} {\bf k}_{n}]),
(a_{LR}\sim 10^{-4})$,
where {\bf s}$_{n}$ , {\bf k}$_{n}$ and {\bf k}$_{\alpha}$ 
are, respectively, the neutron spin, the neutron momentum, and
the momentum of the detected charged particle.
For this reason, the proportional chamber was designed to ensure the
fulfillment of the geometric condition
{\bf s}$_{n} \parallel$ {\bf k}$_{\alpha} \parallel$ {\bf k}$_{n}$.
The detector used in this study consisted of 24 double ionization
chambers positioned in the cylindrical duralumin vessel.
The possible contribution of the left-right asymmetry
to the P-odd effect does not exceed $10^{-8}$.

The relative amplitude of the
fluctuations of the incident neutron flux 
due to temporal changes in
the working parameters of the reactor is about
$10^{-3}$ to $10^{-2}$.
The actual signal variations caused by the nuclear reaction are at
least 30 times smaller ($\sim 10^{-4}$ to $10^{-5}$).
Therefore, it is necessary to compensate the reactor-power fluctuations.
In the forward and backward proportional chambers, the signs of 
the sought effects are opposite to each other. This fact is
used to compensate the reactor-power fluctuations.
In \cite{nim} it is possible to read more about the scheme
of measuring of PV effects for charged particle channels.

An analogous scheme for measuring the $\gamma$-quanta asymmetry differs by
the detector set. Na(Tl) crystals 150 mm long by 150 mm wide with  
photoelectron
multipliers served as detectors. They are housed on both sides of the
sample 70 mm from the neutron beam axis. The integral current
method is used. A  natural boron target is placed in a $^6$LiF box 10 mm thick.
The current signal is taken from the seventh dynode where the linear
dependence between the values of the current and the $\gamma$-quanta flux
occurs.
Special care is
also taken to suppress the magnetic field induced noise.
Low energy $\gamma $-quanta, E$_{\gamma} \sim 500 $ keV, result in an
abrupt (about an order of magnitude) increase of the inherent noise of the
detectors and as a result, a increase of the instrumental error of the
effect under discussion
5 to 10 times that of the statistical error.
We will discuss in short the peculiarities and results of each experiment.

Since the specific energy loss of the $\alpha$-particle in the reaction
$^6$Li(n,$\alpha)^3$H is higher than of the triton,
no precise adjustment of the gas pressure is required.
We use the chamber in the ionization regime without gas enhancement in the
linear range. The total experimental time was about 40 days.
Taking the background measurements into account
the following experimental result for $a^{\alpha}_{PV}$ is obtained
\cite{yd}: $ a^{\alpha}_{PV}=-(6.44 \pm 5.50)\cdot10^{-8}$.

The reaction $^{10}$B(n,$\alpha_{0,1})^7$Li  was studied in the proportional
chamber with an insensitive gas region \cite{nim}.
Two 8.5-day series of measurements were carried out
in the ionization regime: one at an argon pressure of 0.3 atm
(the effect was measured for the mixture of two lines) and the other
at a pressure of 0.78 atm (where according to calculations, only the higher 
energy $\alpha$-group was detected).
The resulting experimental values of the P-odd asymmetry
are as follows \cite{yd}:
$a^{\alpha_1}_{PV}=-(2.5 \pm 1.6)\cdot 10^{-7};
a^{\alpha_0}_{PV}=(3.4 \pm 6.7)\cdot 10^{-7}$.

A relatively novel subject is the measurement of the P-odd asymmetry
$ W^{\gamma}_{PV}(\theta) \sim
1+a_{PV}^{\gamma}({\bf s}_{n}\cdot {\bf k}_ \gamma) $ of the gamma
transition from the first excited state of $ ^7$Li$(1/2 ^-,0.478$ MeV),
populated according to the scheme:
$^{10}$B(n,$\alpha_1)^7$Li$^*(1/2^-,0.478$ MeV) $\rightarrow$ M1(E1)
$\rightarrow ^7$Li$(3/2^-$, g.s.).
In this case the $(n,\alpha_1)$ reaction is used for the
creation of a polarized source of $\gamma$-quanta. Taking into account the
above mentioned instrumental noises only tentative measurements were carried
out. Two two-week series resulted in the value
$a^{\gamma}_{PV}=(6.8 \pm 3.7)\cdot 10^{-7}$,
which is reasonable considering the upper limit of the effect under
investigation because of discussed instrumental errors.
Certainly, the accuracy of this result must be
improved  for the extraction of even the
upper limits of the PV constants. However, the prospect of the increasing 
of the accuracy of the experiment seems to be good. Further instrumental
improvements are the replacement of the photoelectron
multipliers by
semiconductor silicon photodiodes. The latter were explored when the circular
polarization of $np \to d\gamma$ was measured \cite{np}.
The application of
them in this experiment permitted the investigators to compensate the
fluctuations of the reactor almost completely. In the discussed case, the
 Hamamatsu photodiodes with the "dark" current of about
$10^{-9}$A at room temperature seems to be optimal. In this case, the
contribution of the instrumental noise to the total signal is expected to be
negligible. Under these
conditions it is possible to increase the estimated precision
of the measurement by one order of magnitude.

Let us directly consider the theoretical aspects of discussed problem.
In most cases spectra of lightest nuclei do
not contain close PMD and even far distant ones.
So, the P-odd properties of these levels are associated with the admixture of
continuum wave functions of nucleon and cluster channels
or the wave functions of broad resonances
being, in fact, the pieces of the same continuums. To develop the approach,
allowing one to extract the PV constants in this case (notably for
cluster-cluster channels), is
an original theoretical problem which is a rather general and challenging one.

This problem has been partially solved for the first time in Ref. \cite{bdktt},
where the width of the P-odd decay of the $0^+$; T=1 
level of $^6$Li to the $\alpha+d$ continuum is calculated.
In contrast to the discussed case
this case is a typical "one narrow resonance -- one channel" problem \cite{mw},
where the sole wave function of the continuum at the resonance energy
should be known to calculate the decay width.

The discussed processes are essentially different. The first two ones are typical
nuclear reactions characterized not only by resonances but direct mechanism
also. Naturally off-shell effects are essential here and consequently
complete sets of continuum wave functions are required to describe PV in
these cases. Competition between resonance and direct
mechanisms mentioned earlier makes the problem of extracting of PV constants 
very complicated
and we do not discuss it in the present work. The third PV
process is not a nuclear reaction but it is a $\gamma$-transition and
consequently, the theoretical interpretation of it is not so complicated.
Therefore, it is possible to develop a rather well justified approach
to the process.

Let us discuss the P-odd asymmetry of the angular
distribution of $\gamma$-quanta emitted by the polarized sample of first
excited state of $^7$Li  ($1/2^-$, E$^*=0.478$ MeV) as an example of rather
general approach.
The spectrum of the $^7$Li nucleus
(see Ref. \cite{sel}) does not contain any $1/2^+$ or $3/2^+$
discrete levels or narrow resonances (the parity mixing
partners of the initial $1/2^-$ and final $3/2^-$ states).
The lowest continuum of this
7N-system is an $\alpha+t$ continuum beginning at
E$^{\alpha+t}_0=2.47$ MeV.
The thresholds of other channels lie much higher namely 
E$^{^6Li+n}_0=7.25$ MeV, E$^{^6He+p}_0=9.98$ MeV. Such situation is typical
for the discussed nuclear area and so cluster channels are usually the most
important. The existence of broad ($\Gamma\sim 3$ MeV) resonances in
the neutron channel is the subject of controversy. Anyway,
these resonances, even if they really exist, are the pieces of the related
continuums possessing a slightly higher spectral density.
In addition they are removed far away from the levels of our interest.
So, it is rather obvious
that the most essential contribution to the effect under discussion appears
due to the P-odd coupling of the discrete levels $1/2^-$ and $3/2^-$ with the
proper $\alpha+t$ continuum states of positive parity.
So, the expression for the P-odd asymmetry has the form:
$a^{\gamma}_{PV}=
a^{\gamma}_{PV}(excited)+
a^{\gamma}_{PV}(ground)$,
which contains the contributions of the opposite parity components to
both the initial and final states. The formulas for both contributions
are in fact the same.

In comparison with the PMD case the formalism of PV in
the continuum contains
the integration over an infinite energy range of the
$\alpha+t$ continuum. For example, the expression for the wave
function of the excited state of the $^7$Li nucleus $1/2^-$ has the form:
\begin{equation}
{\Psi}_{(1/2)}=\Psi_{(1/2)^-}+
\int\limits_{E_0^{\alpha +t}}^{\infty}
\frac{\langle \Psi_{\alpha+t(1/2)^+}^E| V^{PV}
 |\Psi_{(1/2)^-}\rangle }{E_{(1/2)^-}-E}
 |\Psi_{\alpha +t(1/2)^+}^E \rangle dE.
\label{b}
\end{equation} 
Here, $V^{PV}$ is the PV potential, $\Psi_{\alpha +t(1/2)^+}^E$ is the
wave function of the related positive parity continuum.
The contribution of this state to the discussed effect can be written as:
\begin{equation}
\label{asym}
a^{\gamma}_{PV}(ex.)=A^J_F
 \int\limits_{E_0^{\alpha-t}}^{\infty}
\frac{\langle \Psi_{(3/2)^-}| E1 |\Psi^E_{\alpha+t(1/2)^+} \rangle }
{\langle \Psi_{(3/2)^-}| M1 |\Psi_{(1/2)^-} \rangle }
\frac{\langle \Psi_{\alpha+t(1/2)^+}^E| V^{PV} |\Psi_{(1/2)^-} \rangle }
{E_{(1/2)^-}-E}dE,
\label{d}
\end{equation} 
where the amplitudes of the regular M1 and irregular E1 $\gamma$-transitions
and spin-orbital factor of reaction $A^J_F$
are also the factors in the expression for the PV effect.

The next step is the key to the problem. The
matter is that it is possible to extract the PV constants using a microscopic
two-nucleon expression for the potential $V^{PV}$ in Exp.(\ref{asym})
only. So, it is necessary to write the microscopic 7N wave functions not
only for bound states of $^7$Li but also for the
$\alpha + t$ continuum (or arbitrarily
other nucleon-nucleus or cluster-nucleus continuum for the general
consideration) in the matrix element of this potential.
Dealing with low-laying bound states of lightest nuclei
creates no serious problems.

At the same time, to write the microscopic wave function of the
$\alpha+t$ channel in the form suitable to continue operating with the
microscopic two-nucleon operator $V^{PV}$, we use the method of the
shell-model expansion of the multicluster wave function \cite{ryz}.
As the guide for it the orthogonality condition model (OCM) \cite{saito}
is used. This model is the simplified version of resonating group model
(RGM). The wave function of OCM has the form:
 \begin{equation}
\Psi_{\alpha +t(J)^+}^E = \hat {A}\left \{\Psi_{\alpha} \Psi_t 
\hat{N}^{-\frac{1}{2}}\Phi_l^E(\mbox{\boldmath $\rho$}) \right \}.
\label{f}
\end{equation} 

Here, $\Psi_{\alpha}\Psi_t$ are the internal wave functions of the
alpha-particle and the triton, respectively, 
$\Phi_l^E(\mbox{\boldmath $\rho$})$ is
the function of relative
motion in the $\alpha+t$ continuum, $\hat A$ is the antisymmetrizer and the
operator $\hat N$ is the exchange kernel of RGM. As it is shown in
\cite{Yamani} one can expand the continuum wave function
$\Phi_l^E(\mbox{\boldmath $\rho$})$  in terms of the oscillator wave functions
$\Phi_{nl}(\mbox{\boldmath $\rho$})$.
Due to this, for the two-body continuum wave functions one can obtain:
 \begin{equation}
\Psi_{\alpha +t(J)^+}^E=\sum_{n}C_n(E)
\varepsilon_n^{-1/2}\hat{A}\{\Psi_{\alpha} \Psi_t \Phi_{nl}
(\mbox{\boldmath $\rho$})\},
\label{i}
\end{equation} 
where $C_n(E)$ are the coefficients of the oscillator expansion of the
function $\Phi_l^E(\mbox{\boldmath $\rho$})$  and $\varepsilon_n$
are the eigenvalues of the exchange kernel.

Further transformation of the continuum wave functions
$\Psi_{\alpha +t(J)^+}^E$
is the conversion from the seven-nucleon two-cluster oscillator wave functions
contained in the right-hand side of Exp. (\ref{i}) to the
translationally-invariant shell model (TISM) wave
functions. It is important to note  that the first term of Exp. (\ref{i})
which does not vanish by the antisymmetrizer $(n=4)$ coincides with the
respective TISM wave function. But it is this
function that is the sole component, contributing to the matrix element
of the E1 transition because the functions of the low lying levels of $^7$Li
are characterized by the principal quantum number $n=3$ and the operator
of the E1 transition changes that number by unity.

Inserting the expressions for the wave functions into the formula (\ref{d})
and taking into account the above mentioned circumstances we obtain:
 \begin{equation}
a^{\gamma}_{PV}(ex.)=A^J_F
\sum_n\frac{\langle \Psi_{(3/2)^-}|E1|{\Psi}_{4,(1/2)^+} \rangle
\langle {\Psi}_{n,(1/2)^+}|V^{PV}|\Psi_{(1/2)^-} \rangle }
{\langle \Psi_{(3/2)^-}| M1 |\Psi_{(1/2)^-} \rangle }
\int\limits_{E_0^{\alpha -t}}^{\infty}
\frac{C^*_n(E)C_4(E)}
{E_{(1/2)^-}-E}dE.
\label{j}
\end{equation} 
Here we do not present the cumbersome transformation of the wave function
${\Psi}_{n,(1/2)^+}=\varepsilon_n^{-1/2}
\hat{A}\{\Psi_{\alpha}\Psi_t\Phi_{nl}
(\mbox{\boldmath $\rho$})\}$ to the linear combination of TISM 
wave functions for brevity.
The wave function
$\Phi_l^E(\mbox{\boldmath $\rho$})$
is the solution of the Schrodinger equation
containing the nuclear and Coulomb terms.
The Woods-Saxon form of the cluster-cluster interaction
with the parameters $V_0=87$ MeV, $\rho_0=1.8$ fm and
$a=0.7$ fm proposed in \cite{neud} to fit the phase shifts of
$\alpha-t$ scattering is used.

So, the scheme which is deduced is in fact an universal theoretical approach
to the solution of the problem of the P-odd mixing of a discrete level and
arbitrary two body (nucleon-nucleus or cluster-nucleus) 
continuum.

For the particular case discussed, our main goal was to show that the
scale of the PV effect here is not small in comparison with the experimental
precision. In the light of this the precision of the used model seems to be
quite sufficient.

The final formula for the P-odd asymmetry of the transition under
investigation, written in terms of PV constants, takes the form:

 \begin{equation}
a^\gamma_{PV} = - 0.078\cdot h_{\pi}
+ 0.028\cdot h_{\rho}^0 + 0.010\cdot h_{\rho}^1
+ 0.015\cdot h_{\omega}^0 + 0.014\cdot h_{\omega}^1.
\label{a10}
\end{equation} 
It takes the value $a^\gamma_{PV} = -4.17\cdot 10^{-8}$ if one uses
the Dubovik-Zenkin \cite{b2} version of PV-constants  and
$a^\gamma_{PV} = -7.24\cdot10^{-8}$ if one uses
DDH "best values" \cite{b1}. The contribution of
neutral currents to the discussed process is
equal to: $a^\gamma_{NC} =  -1.28\cdot 10^{-8}$ (30.6\% ) for
PV-constants from \cite{b2} and
$a^\gamma_{NC} = -3.76\cdot 10^{-8}$
(51.9\%) for "best values" from \cite{b1}.
So predicted values of the effect turn out to be measurable by the
discussed method .

To demonstrate qualitatively the scale of continuum PV effects it is
convenient to define the value of
$\bigtriangleup$ E$^{eff}$ as the equivalent energy distance between a level
and its nonexisting PV partner producing the
effect of continuum mixing. The use of $n=4$ term matrix elements brings the 
results: $\bigtriangleup$ E$^{eff}=$17.8 MeV 
and 21.2 MeV for excited $1/2^-$ and ground $3/2^-$
states, respectively. The effect of
continuum mixing is equal to that of a discrete
level spaced an approximately $\hbar\omega$ energy distance
apart from the top of the potential barrier although for convergency
of the results it is necessary to take
into account the influence of wide range of
continuum wave functions up to the energy value close to 300 MeV.
So the effect of the continuum with a
rather close threshold and a low barrier has the scale similar to that of
a nonenhanced process.

The effect of
continuum turns out to be tolerant to manifold variations of the
input namely the model of cluster-cluster interaction (
if realistic versions of it are used), 
cluster wave
functions etc. This property appears, of course, if any narrow resonance is
absent in a continuum. In this case the integral with respect to E 
in (\ref{asym}) is a smooth function of the upper limit.
That's why the related
values of $\bigtriangleup$ E$^{eff}$ are notably independent on the model of
interaction. In this
respect, the effect under discussion is even more preferable for extracting
the PV constants than the PV effect
caused by PMD being more tolerant to the nuclear peculiarities. Naturally
that is true if PMD are absent in the spectra of
clusters.

The versatility of the method of shell model
expansion, used in the present work, 
allows one to incorporate rather easily more precise models of clusters and
cluster-cluster interaction. For example it is possible to take into account
another channels, to use one- or multi-channel RGM in place of OCM.
The extension of the approach to the PV effect in direct reactions
is also not hopeless.

So the discussed measurements deserve serious theoretical
support.

In conclusion let us note that the present work demonstrates
the possibilities of modern experiments to measure
P-odd effects in above mentioned nonenhanced nuclear processes
i.e. to achieve the precision of the
order of several $1 \cdot 10^{-8}$ using the peculiarities of lightest nuclei.
Good prospects for theory in the discussed area are
also demonstrated. The result of calculation of the P-odd
asymmetry of the 478-keV $\gamma$-transition in the polarized $^7$Li 
nucleus demonstrates that the above mentioned precision is 
satisfactory for the extracting of
the constants of the PV NN potential. And what is more, the contribution of
neutral currents is large enough to be measured.
It is shown that the existence of PMD in the discussed systems 
is not a strict condition for a successful investigation of PV
effects. So, new prospects for the study of PV constants
in the reported area of nuclear masses, in general, and in the discussed
reactions, in particular, are seen. The $\gamma$-transition between two
lowest levels in a polarized $^7$Li nucleus is believed to be the most
promising for these purposes. Finally let us conclude that extended 
experimental investigations of the last-named process
using a high-intensity thermal neutron source seems us to be very desirable, 
so the present work may be considered as a preliminary proposal for them.

The work supported by RFBR \mbox{grants $N^{\underline{o}}$
00-02-16707} (experimental part) \mbox{and $N^{\underline{o}}$
00-02-16683} (theoretical part).

\end{document}